\documentstyle[twoside,fleqn,espcrc2]{article}
\input epsf

\title{Evidence of Strong Correlation between Instanton and QCD-monopole
on SU(2) Lattice}

\author{\underline{H.~Suganuma}\address{Research Center for Nuclear Physics 
(RCNP), Osaka University, Mihogaoka 10-1, Ibaraki 567, Japan}, 
A.~Tanaka${\rm ^a}$, S.~Sasaki${\rm ^a}$
and
O.~Miyamura\address{Department of Physics, Hiroshima University,
Kagamiyama 1-3, Higashi-Hiroshima 739, Japan}
}
       
\begin{document}

\begin{abstract}
The correlation between instantons and QCD-monopoles is studied 
both in the lattice gauge theory and in the continuum theory.
An analytical study in the Polyakov-like gauge,
where $A_4(x)$ is diagonalized, shows that the QCD-monopole trajectory 
penetrates the center of each instanton, and becomes complicated 
in the multi-instanton system. Using the SU(2) lattice with $16^4$,
the instanton number is measured in the singular (monopole-dominating) 
and regular (photon-dominating) parts, respectively.
The monopole dominance for the topological charge is found both
in the maximally abelian gauge and in the Polyakov gauge.
\end{abstract}

\maketitle

\section{Introduction}

QCD is reduced to an abelian gauge theory 
with magnetic monopoles (QCD-monopoles) by the abelian gauge fixing 
through the diagonalization of a gauge dependent variable $X(x)$ 
\cite{thooft}.
The QCD-monopole appears from the hedgehog configuration on $X(x)$ 
corresponding to the nontrivial homotopy group 
$\pi _2({\rm SU}(N_c)/{\rm U(1)}^{N_c-1})=Z_\infty ^{N_c-1}$, 
and its condensation  
plays an essential role to the nonperturbative QCD \cite{kronfeld}.
The instanton is also an important topological object 
relating to U$_{\rm A}$(1) anomaly, 
and appears in the Euclidean 4-space 
corresponding to $\pi _3({\rm SU}(N_c))=Z_\infty $. 
We study the correlation between instantons and QCD-monopoles 
both in the lattice theory \cite{origuchi}
and in the analytical framework \cite{suganuma}.

Recent lattice studies \cite{kronfeld}
indicate the abelian dominance for the nonperturbative quantities 
in the maximally abelian (MA) gauge and/or in the Polyakov gauge. 
If the system is completely described only by the abelian field, 
the instanton would lose the topological basis 
for its existence, and therefore it seems unable to 
survive in the abelian manifold. 
However, even in the abelian gauge, nonabelian components remain 
relatively large around the QCD-monopoles, which are nothing 
but the topological defects, so that instantons 
are expected to survive only around the QCD-monopole trajectories 
in the abelian-dominant system.
The close relation between instantons and QCD-monopoles are thus 
suggested from the topological consideration.

\section{Analytic Calculation}

First, we demonstrate a close relation between instantons 
and QCD-monopoles within the continuum theory \cite{suganuma}. 
Using an ambiguity on $X(x)$ in the abelian gauge fixing,
we choose $X(x)=A_4(x)$ to this end.
This abelian gauge diagonalizing $A_4(x)$ will be called as the
Polyakov-like gauge, where the Polyakov loop $P(x)$ is also diagonal.
Since $A_4(x)$ takes a hedgehog configuration around each
instanton, the QCD-monopole trajectory should pass 
through the center of instantons
inevitably in the Polyakov-like gauge.
We show this relation in the SU(2) gauge theory below.

Using the 't~Hooft symbol ${\bar \eta}^{a\mu \nu }$, 
the multi-instanton solution is written as 
{\setlength{\mathindent}{3pt}
\begin{equation}
A^{\mu} (x) = i{\bar \eta}^{a\mu \nu} {\tau^a \over 2} \partial^\nu 
\ln \left( 1+\sum_k {a_k^2 \over |x-x_k|^2} \right), 
\label{ONE}
\end{equation}
}
\noindent
where $x_k^\mu  \equiv ({\bf x}_k,t_k)$ and $a_k$ denote the center 
coordinate and the size of $k$-th instanton, respectively. 
Near the center of $k$-th instanton, 
$A_4(x)$ takes a hedgehog configuration, 
\begin{equation}
A_4(x) \simeq i {\tau ^a ({\bf x}-{\bf x}_k)^a \over |x-x_k|^2}.
\label{TWO}
\end{equation}
In the Polyakov-like gauge, 
$A_4(x)$ is diagonalized by a singular gauge transformation, 
which leads to the QCD-monopole trajectory on $A_4(x)=0$: 
${\bf x} \simeq {\bf x}_k$. 
Thus, the center of each instanton is penetrated 
by a QCD-monopole trajectory with the temporal direction 
in the Polyakov-like gauge \cite{suganuma}. 
In other words, instantons only live along the 
QCD-monopole trajectories. 

Here, we refer the magnetic charge of the QCD-monopole.
In general, the abelian gauge fixing consists of two sequential 
procedures. One is the diagonalization of $X(x)$: $X(x)\rightarrow X_d(x)$. 
The other is the ordering of the diagonal elements of $X_d(x)$, 
e.g., $X_d^1 \ge X_d^2 \ge ... \ge X_d^{N_c}$.
The gauge group SU($N_c$) is reduced to U(1)$^{N_c-1}\times P_{N_c}$
by the diagonalization of $X(x)$, and is reduced to U(1)$^{N_c-1}$
by the ordering condition on $X_d(x)$.
The magnetic charge of the QCD-monopole is closely related to 
the ordering condition in the diagonalization in the abelian 
gauge fixing. 
For instance, in the SU(2) case, 
the hedgehog configuration as $X(x) \sim ({\bf x} \cdot \tau )$ and 
the anti-hedgehog one as $X(x) \sim -({\bf x} \cdot\tau )$ 
provide a QCD-monopole with an opposite magnetic charge respectively, 
because they are connected by the additional gauge transformation 
$\Omega =\exp\{i\pi (\tau ^1\cos\phi +\tau ^2\sin\phi )\}$, which interchanges 
the diagonal elements of $X_d(x)$ and leads a minus sign in the 
U(1)$_3$ gauge field.

For the single-instanton system, 
the QCD-monopole trajectory $x^\mu \equiv({\bf x},t)$ is simply 
given by ${\bf x}={\bf x}_1$ $(-\infty <t<\infty )$ at the classical level.

For the two-instanton system, two instanton centers 
can be located on the $zt$-plane without loss of 
generality, so that one can set $x_1=y_1=x_2=y_2=0$. 
Owing to the symmetry of the system, 
QCD-monopoles only appear on the $zt$-plane, and hence 
one has only to examine $A_4(x)$ on the $zt$-plane by setting $x=y=0$. 
In this case, $A_4(x)$ in Eq.(\ref{ONE}) is already 
diagonalized on the $zt$-plane: $A_4(x)=A_4^3(z,t)\tau ^3$. 
Therefore, the QCD-monopole trajectory $x^\mu =(x,y,z,t)$ 
is simply given by $A_4^3(z,t)=0$ and $x=y=0$. 
However, the QCD-monopole trajectories are rather complicated 
even at the classical level in the two-instanton system. 
According to the parameters $x_k, a_k$ ($k=1,2$), 
the QCD-monopole trajectory has a loop or a folded structure 
as shown in Fig.1 (a) or (b), respectively. 
Here, the QCD-monopole trajectories originating from instantons 
are very unstable against a small fluctuation relating to 
the location or the size of instantons.

The QCD-monopole trajectory tends to be highly complicated 
and unstable in the multi-instanton system even at the classical level, 
and the topology of the trajectory is often changed due to a small 
fluctuation of instantons.
In addition, the quantum fluctuation would make it more complicated 
and more unstable, which leads to appearance of a long complicated 
trajectory as a result.
Thus, instantons may contribute to promote monopole condensation, 
which is signaled by a long complicated monopole loop in the 
lattice QCD simulation \cite{kronfeld}.

We also study the thermal instanton system in the Polyakov-like gauge.
At high temperature, QCD-monopole trajectories 
tend to be reduced to simple straight lines penetrating instantons 
in the temporal direction, 
which may corresponds to the deconfinement phase transition through 
the vanishing of QCD-monopole condensation.

\begin{figure}[bh]
\centerline{\epsfbox{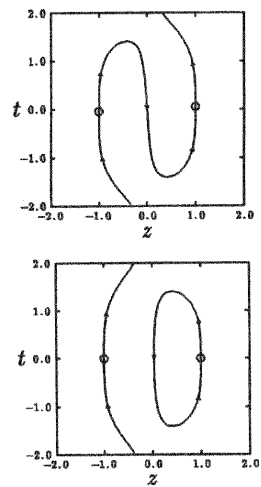}}
\caption{Examples of the QCD-monopole trajectory in the two-instanton system
with (a) $(z_1,t_1)=-(z_2,t_2)=(1,0.05)$, $a_1 = a_2$ and (b)
$(z_1,t_1)=-(z_2,t_2)=(1,0)$, $a_2 = 1.1 a_1$.}
\end{figure}

\noindent
For the thermal two-instanton system, 
the topology of the QCD-monopole trajectory is drastically changed 
at $T_c \simeq 0.6 d^{-1}$, where $d$ is the distance between the 
two instantons. If one adopts $d \sim 1{\rm fm}$ 
as a typical mean distance between instantons, such a topological 
change occurs at $T_c \sim 120 {\rm MeV}$.

\section{Instanton and Monopole on Lattice}

We study the correlation between instantons and QCD-monopoles 
in the maximally abelian (MA) gauge and in the Polyakov gauge 
using the SU(2) lattice with  $16^4$ and $\beta =2.4$.
All measurements are done every 500 sweeps after a 
thermalization of 1000 sweeps using the heat-bath algorithm. 
After generating the gauge configurations, 
we examine the monopole dominance 
\cite{kronfeld,origuchi} 
for the topological charge using the following procedure.

\vspace{6pt}
\noindent
1)\hspace{2pt}The abelian gauge fixing is done by diagonalizing 
$R(s)=\sum_{\mu } U_\mu (s)\tau ^3U^{-1}_\mu (s)$ in the MA gauge, 
or the Polyakov loop $P(s)$ in the Polyakov gauge. 

\vspace{6pt}
\noindent
2)\hspace{2pt}The SU(2) link variable $U_\mu (s)$ is factorized as 
$U_\mu (s)=M_\mu (s)u_\mu (s)$ with the `off-diagonal' factor 
$M_\mu (s) \equiv \exp\{i\tau ^1C^1_\mu (s)+i\tau ^2C^2_\mu (s)\}$ 
and the abelian link variable $u_\mu (s)=\exp\{i\tau _3\theta _\mu (s)\}$. 

\vspace{6pt}
\noindent
3)\hspace{2pt}The abelian field strength 
$\theta _{\mu \nu } \equiv \partial_\mu \theta _\nu -\partial_\nu \theta _\mu $ 
is decomposed as 
$\theta _{\mu \nu }(s)=\bar \theta _{\mu \nu }(s)+2\pi M_{\mu \nu }(s)$ 
with $-\pi <\bar \theta _{\mu \nu }(s)<\pi $ and $M_{\mu \nu }(s) \in {\bf Z}$ 
\cite{DGT}.
Here, $\bar \theta _{\mu \nu }(s)$ and $2\pi M_{\mu \nu }(s)$ correspond to 
the regular photon and the Dirac string, respectively

\vspace{6pt}
\noindent
4)\hspace{2pt}The U(1) gauge field $\theta _\mu (s)$ is decomposed 
as $\theta _\mu (s)=\theta ^{Ph}_\mu (s)+\theta ^{Ds}_\mu (s)$ with 
a regular part $\theta ^{Ph}_\mu (s)$ and a singular part 
$\theta ^{Ds}_\mu (s)$, which are obtained from 
$\bar \theta _{\mu \nu }(s)$ and $2\pi M_{\mu \nu }(s)$, respectively, 
using the lattice Coulomb propagator in the Landau gauge 
\cite{kronfeld,DGT}.
The singular part carries almost the same amount of 
magnetic current as the original U(1) field, whereas 
it scarcely carries the electric current. 
The situation is just the opposite in the regular part.
For this reason, we regard the singular part as `monopole-dominating', 
and the regular part as `photon-dominating' \cite{origuchi}. 

\vspace{6pt}
\noindent
5)~The corresponding SU(2) variables are reconstructed 
from $\theta ^{Ph}_\mu (s)$ and $\theta ^{Ds}_\mu (s)$ 
by multiplying the off-diagonal factor $M_\mu (s)$ : 
$U^{Ph}_\mu (s) = M_\mu (s)\exp\{i\tau _3\theta _\mu ^{Ph}(s)\}$ 
and $U^{Ds}_\mu (s) = M_\mu (s)\exp\{i\tau _3\theta _\mu ^{Ds}(s)\}$.

\vspace{6pt}
\noindent
6)\hspace{2pt}The topological charge 
$Q= {\int {d^4x \over 16\pi ^2}} {\rm tr} (G_{\mu \nu }\tilde G_{\mu \nu })$, 
the integral of the absolute value of the topological density 
$I_Q \equiv \int {d^4x \over 16\pi ^2} |{\rm tr}(G_{\mu \nu }\tilde G_{\mu \nu })|$, 
and the action $S$ 
are calculated by using $U_\mu (s)$, $U^{Ph}_\mu (s)$ and $U^{Ds}_\mu (s)$.
Then, three sets of quantities are obtained,
$\{Q({\rm SU(2)})$, $I_Q({\rm SU(2)})$, $S({\rm SU(2)})\}$ 
for the full SU(2) variable,
$\{Q({\rm Ph})$,    $I_Q({\rm Ph})$,    $S({\rm Ph})\}$
for the regular  part, and 
$\{Q({\rm Ds})$,    $I_Q({\rm Ds})$,    $S({\rm Ds})\}$
for the singular part.
Here, $I_Q$ has been introduced to get information on 
the instanton and anti-instanton pair.

\vspace{6pt}
\noindent
7)\hspace{2pt}The correlations among these quantities are 
examined using the Cabibbo-Marinari cooling method.

\vspace{6pt}
\begin{figure}[bh]
\centerline{\epsfbox{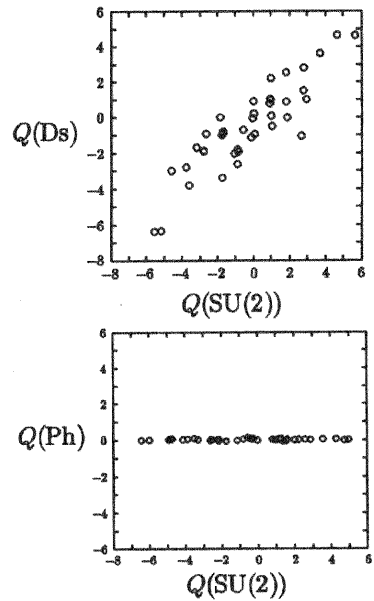}}
\caption{Correlations between (a) $Q({\rm Ds})$ and $Q({\rm SU(2)})$
at 80 cooling sweeps, (b) $Q({\rm Ph})$ and $Q({\rm SU(2)})$ 
at 10 cooling sweeps.}
\end{figure}

\indent
We prepare 40 samples for the MA gauge and the Polyakov gauge, respectively.
These simulations have been performed on the Intel Paragon XP/S(56node) at 
the Institute for Numerical Simulations and Applied Mathematics of Hiroshima
University. 
Since quite similar results have been obtained in the MA gauge 
\cite{origuchi} and in the Polyakov gauge, only latter case is shown. 

Fig.2 shows the correlation among $Q({\rm SU(2)})$, 
$Q({\rm Ds})$ and $Q({\rm Ph})$ after some cooling sweeps 
in the Polyakov gauge. 
A strong correlation is found between 
$Q({\rm SU(2)})$ and $Q({\rm Ds})$, which is defined in singular 
(monopole-dominating) part. Such a strong correlation 
remains even at 80 cooling sweeps.
On the other hand, $Q({\rm Ph})$ quickly vanishes only by several 
cooling sweeps, and no correlation is seen between $Q({\rm Ph})$ 
and $Q({\rm SU(2)})$.

\begin{figure}[bh]
\centerline{\epsfbox{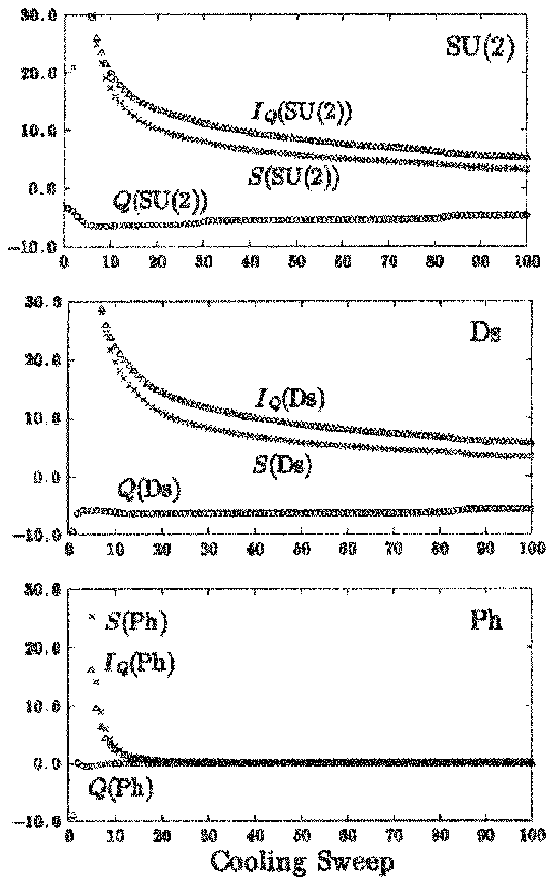}}
\caption{Cooling curves for 
(a) $Q({\rm SU(2)})$, $I_Q({\rm SU(2)})$, $S({\rm SU(2)})$,
(b) $Q({\rm Ds})$, $I_Q({\rm Ds})$, $S({\rm Ds})$,
(c) $Q({\rm Ph})$, $I_Q({\rm Ph})$, $S({\rm Ph})$.}
\end{figure}

We show in Fig.3 the cooling curves for $Q$, $I_Q$ and $S$ 
in a typical example with $Q({\rm SU(2)})\neq0$ 
in the Polyakov gauge. 
Similar to the full SU(2) case, 
$Q({\rm Ds})$, $I_Q({\rm Ds})$ and $S({\rm Ds})$ in the singular 
(monopole-dominating) part tends to remain finite 
during the cooling process.
On the other hand, $Q({\rm Ph})$, $I_Q({\rm Ph})$ 
and $S({\rm Ph})$ in the regular part quickly vanish by 
only less than 10 cooling sweeps. 
Therefore, instantons seem unable to live in the regular 
(photon-dominating) part, but only survive in the 
singular (monopole-dominating) part in the abelian gauges. 

The cooling curves for $Q$, $I_Q$ and $S$ are examined in the case
with $Q({\rm SU(2)})=0$. 
Similar to the full SU(2) case, 
$I_Q({\rm Ds})$ and $S({\rm Ds})$ decrease slowly and 
remain finite even at 70 cooling sweeps,  
which means the existence of the instanton and 
anti-instanton pair in the singular part. 
On the other hand, $I_Q({\rm Ph})$ and $S({\rm Ph})$ quickly
vanish, which indicates the absence of such a topological pair
excitation in the regular part.

In conclusion, the monopole dominance for the topological charge 
is found both in the MA gauge and in the Polyakov gauge. 
In particular, instantons would survive only in the singular 
(monopole-dominating) part in the abelian gauges,
which agrees with the result in our previous analytical
study. 
The monopole dominance for the U$_{\rm A}$(1) anomaly is also expected.

\end{document}